\documentclass[sigconf, anonymous=false]{acmart}

\usepackage{booktabs} 

\usepackage{graphicx}
\usepackage{caption} 
\usepackage{subcaption} 

\usepackage{amsfonts}       
\usepackage{microtype}      
\usepackage{url}
\usepackage{verbatim} 
\usepackage{multirow}
\usepackage[linesnumbered,ruled]{algorithm2e}
\usepackage{xspace}
\usepackage{epsfig}
\usepackage{amsthm}
\usepackage{amssymb}
\usepackage{xr}
\usepackage{enumitem}
\usepackage{bm}
\usepackage{balance}

\newcommand{\note}[1]{{{\textcolor{blue}{[note: #1]}}}}

\newcommand{\rex}[1]{{{\textcolor{magenta}{[Rex: #1]}}}}
\newcommand{\will}[1]{{{\textcolor{teal}{[Will: #1]}}}}

\newcommand{\xhdr}[1]{\vspace{0.75mm}\noindent{{\bf #1.}}}
\newcommand{\cut}[1]{}

\newcommand{\name}{PinSage\xspace}
\newcommand{\framework}{GCN\xspace}

\newcommand{\mb}{\mathbf}

\newcommand{\R}{\mathbb{R}}
\newcommand{\eg}{\emph{e.g.}}
\newcommand{\ie}{\emph{i.e.}}

\newcommand*{\fnref}[1]{\textsuperscript{\ref{#1}}}

\setcopyright{rightsretained}

\copyrightyear{2018} 
\acmYear{2018} 
\setcopyright{acmcopyright}
\acmConference[KDD '18]{The 24th ACM SIGKDD International Conference on Knowledge Discovery \& Data Mining}{August 19--23, 2018}{London, United Kingdom}
\acmBooktitle{KDD '18: The 24th ACM SIGKDD International Conference on Knowledge Discovery \& Data Mining, August 19--23, 2018, London, United Kingdom}
\acmPrice{15.00}
\acmDOI{10.1145/3219819.3219890}
\acmISBN{978-1-4503-5552-0/18/08}
\begin{document}
\title{Graph Convolutional Neural Networks for Web-Scale Recommender Systems}

\author{Rex Ying$^{*\dagger}$, Ruining He$^{*}$, Kaifeng Chen$^{*\dagger}$, Pong Eksombatchai$^{*}$, \\William L. Hamilton$^{{\dagger}}$, Jure Leskovec$^{*\dagger}$}
\affiliation{%
  \institution{$^*$Pinterest, $^\dagger$Stanford University}
}
\email{{rhe, kaifengchen, pong}@pinterest.com, {rexying, wleif,jure}@stanford.edu}

\cut{
\author{Rex Ying}
\authornote{}
\orcid{0001-8131-6311}
\affiliation{%
  \institution{Stanford University}
  \streetaddress{}
  \city{Stanford} 
  \state{CA} 
  \postcode{94305}
}
\email{rexying@stanford.edu}

\author{Ruining He}
\authornote{}
\orcid{}
\affiliation{%
  \institution{Pinterest, Inc.}
  \streetaddress{}
  \city{San Francisco} 
  \state{CA} 
  \postcode{94107}
}
\email{rhe@pinterest.com}

\author{Kaifeng Chen}
\authornote{}
\orcid{}
\affiliation{%
  \institution{Stanford University}
  \streetaddress{}
  \city{Stanford} 
  \state{CA} 
  \postcode{94305}
}
\email{kaifengchen@pinterest.com}

\author{Pong Eksombatchai}
\authornote{}
\orcid{}
\affiliation{%
  \institution{Pinterest, Inc.}
  \streetaddress{}
  \city{San Francisco} 
  \state{CA} 
  \postcode{94107}
}
\email{pong@pinterest.com}

\author{William L. Hamilton}
\authornote{}
\orcid{}
\affiliation{%
  \institution{Stanford University}
  \streetaddress{}
  \city{Stanford} 
  \state{CA} 
  \postcode{94305}
}
\email{wleif@stanford.edu}

\author{Jure Leskovec}
\authornote{}
\orcid{}
\affiliation{%
  \institution{Stanford University}
  \streetaddress{}
  \city{Stanford} 
  \state{CA} 
  \postcode{94305}
}
\email{jure@stanford.edu}
}
\renewcommand{\shortauthors}{R. Ying et al.}
\begin{abstract}


Recent advancements in deep neural networks for graph-structured data have led to state-of-the-art performance on recommender system benchmarks. However, making these methods practical and scalable to web-scale recommendation tasks with billions of items and hundreds of millions of users remains a challenge.

Here we describe a large-scale deep recommendation engine that we developed and deployed at Pinterest. 
We develop a data-efficient Graph Convolutional Network (GCN) algorithm \name, which
combines efficient random walks and graph convolutions to generate embeddings of nodes 
(i.e., items) that incorporate both graph structure as well as node feature information. 
Compared to prior GCN approaches, we develop a novel method based on highly efficient
random walks to structure the convolutions and design a
novel training strategy that relies on harder-and-harder training
examples to improve robustness and convergence of the model. 

We deploy \name at Pinterest and train it on 7.5 billion examples on a graph with 3 billion nodes representing pins and boards, and 18 billion edges. 
According to offline metrics, user studies and A/B tests, \name generates higher-quality recommendations than comparable deep learning and graph-based alternatives.
To our knowledge, this is the largest application of deep graph embeddings to date and paves the way for a new generation of web-scale recommender systems based on graph convolutional architectures.


\end{abstract}

%
%
%
%

\maketitle

\cut{
-- Neural networks revolution in rec sys.
-- However most neural networks are only content based and they cannot account for relational information (collaborative filtering) and this way fuse content plus relation. 
-- On the other hand, recent deep convolutional architectures for graph-structured data have trouble scaling to anything more than few million nodes.
-- Here we presen S-GCN where we extend GCN [x] and make it scalable.

Our innovations are:
-- stochastic s-gcn, which means we keep the nbrghood size constant
-- our approach combines content features (visual, text) with CF-style bipartite graph information (user-movie)
-- mean pooling, loss function makes a huge difference (crucial)
-- we use curriculum learning for data-efficiency and speed up the learning (explain why this speeds up the learning). scale the learning rate.
-- map reduce inference scheme
}

\section{Introduction}
\label{sec:intro}

\begin{figure*}[t!]
\centering
\includegraphics[width=0.80\textwidth]{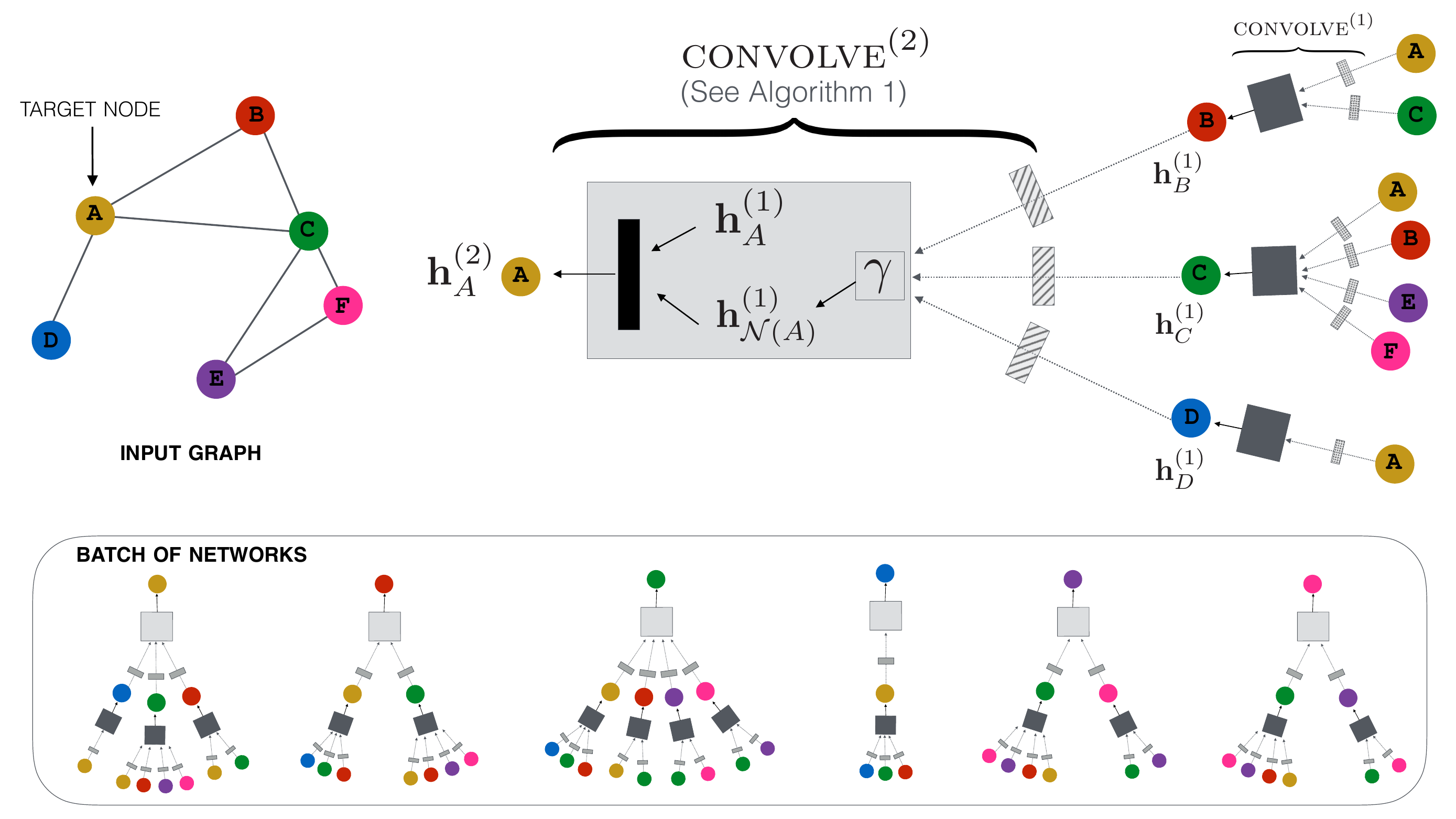}
\vspace{-2mm}
\caption{Overview of our model architecture using depth-2 convolutions (best viewed in color). Left: A small example input graph. Right: The 2-layer neural network that computes the embedding $\mb{h}_A^{(2)}$ of node $A$ using the previous-layer representation, $\mb{h}_A^{(1)}$, of node $A$ and that of its neighborhood $\mathcal{N}(A)$ (nodes $B, C, D$). 
(However, the notion of neighborhood is general and not all neighbors need to be included (Section \ref{sec:arch}).)
Bottom: The neural networks that compute embeddings of each node of the input graph. While neural networks differ from node to node they all share the same set of parameters (\textit{i.e.}, the parameters of the $\textsc{convolve}^{(1)}$ and $\textsc{convolve}^{(2)}$ functions; Algorithm 1). Boxes with the same shading patterns share parameters; $\gamma$ denotes an importance pooling function; and thin rectangular boxes denote densely-connected multi-layer neural networks. 
} 
\vspace{-3mm}
\label{fig:model}
\end{figure*}

Deep learning methods have an increasingly critical role in recommender system applications, being used to learn useful low-dimensional embeddings of images, text, and even individual users \cite{covington2016deep,van2013deep}.
The representations learned using deep models can be used to complement, or even replace, traditional recommendation algorithms like collaborative filtering. 
and these learned representations have high utility because they can be re-used in various recommendation tasks. 
For example, item embeddings learned using a deep model can be used for item-item recommendation and also to recommended themed collections (\eg, playlists, or ``feed'' content).

Recent years have seen significant developments in this space---especially the development of new deep learning methods that are capable of learning on graph-structured data, which is fundamental for recommendation applications (\eg, to exploit user-to-item interaction graphs as well as social graphs)~\cite{berg2017graph,bronstein2017geometric,hamilton2017representation,kipf2016semi,monti2017geometric,you2018graphrnn}.

Most prominent among these recent advancements is the success of deep learning architectures known as Graph Convolutional Networks (GCNs)~\cite{berg2017graph,hamilton2017representation,kipf2016semi,monti2017geometric}. 
The core idea behind GCNs is to learn how to iteratively aggregate feature information from local graph neighborhoods using neural networks (Figure~\ref{fig:model}).
Here a single ``convolution'' operation transforms and aggregates feature information from a node's one-hop graph neighborhood, and by stacking multiple such convolutions information can be propagated across far reaches of a graph. 
Unlike purely content-based deep models (\eg, recurrent neural networks~\cite{bansal2016ask}), GCNs leverage both content information as well as graph structure. GCN-based methods have set a new standard on countless recommender system benchmarks (see \cite{hamilton2017representation} for a survey). However, these gains on benchmark tasks have yet to be translated to gains in real-world production environments. 

The main challenge is to scale both the training as well as inference of GCN-based node embeddings to graphs with billions of nodes and tens of billions of edges. Scaling up GCNs is difficult because many of the core assumptions underlying their design are violated when working in a big data environment. For example, all existing GCN-based recommender systems require operating on the full graph Laplacian during training---an assumption that is infeasible when the underlying graph has billions of nodes and whose structure is constantly evolving.

\vspace{10pt}
\xhdr{Present work}
Here we present a highly-scalable GCN framework that we have developed and deployed in production at Pinterest.
Our framework, a random-walk-based GCN named \name, operates on a massive graph with 3 billion nodes and 18 billion edges---a graph that is $10,000\times$ larger than typical applications of GCNs. 
\name leverages several key insights to drastically improve the scalability of GCNs:
\begin{itemize}[leftmargin=10pt, topsep=3pt]
\item
	{\bf On-the-fly convolutions:} Traditional GCN algorithms perform graph convolutions by multiplying feature matrices by powers of the full graph Laplacian.
	In contrast, our \name algorithm performs efficient, localized convolutions by sampling the neighborhood around a node and dynamically constructing a computation graph from this sampled neighborhood. 
	These dynamically constructed computation graphs (Fig.~\ref{fig:model}) specify how to perform a localized convolution around a particular node, and alleviate the need to operate on the entire graph during training.
\item
	{\bf Producer-consumer minibatch construction:} We develop a producer-consumer architecture for constructing minibatches that ensures maximal GPU utilization during model training. A large-memory, CPU-bound producer efficiently samples node network neighborhoods and fetches the necessary features to define local convolutions, while a GPU-bound TensorFlow model consumes these pre-defined computation graphs to efficiently run stochastic gradient decent.
\item
	{\bf Efficient MapReduce inference:} Given a fully-trained GCN model, we design an efficient MapReduce pipeline that can distribute the trained model to generate embeddings for billions of nodes, while minimizing repeated computations. 
\end{itemize} 
In addition to these fundamental advancements in scalability, we also introduce new training techniques and algorithmic innovations.
These innovations improve the quality of the representations learned by \name, leading significant performance gains in downstream recommender system tasks:
\begin{itemize}[leftmargin=10pt, topsep=1pt]
\item 
	{\bf Constructing convolutions via random walks:} Taking full neighborhoods of nodes to perform convolutions (Fig.~\ref{fig:model}) would result in huge computation graphs, so we resort to sampling. However, random sampling is suboptimal, and we develop a new technique using short random walks to sample the computation graph. An additional benefit is that each node now has an importance score, which we use in the pooling/aggregation step.
\item
	{\bf Importance pooling:} A core component of graph convolutions is the aggregation of feature information from local neighborhoods in the graph.
	We introduce a method to weigh the importance of node features in this aggregation based upon random-walk similarity measures, leading to a $46\%$ performance gain in offline evaluation metrics. 
\item
	{\bf Curriculum training:} We design a curriculum training scheme, where the algorithm is fed harder-and-harder examples during training, resulting in a $12\%$ performance gain. 
\end{itemize} 

We have deployed \name for a variety of recommendation tasks at Pinterest, a popular content discovery and curation application where users interact with {\em pins}, which are visual bookmarks to online content (\textit{e.g.}, recipes they want to cook, or clothes they want to purchase). 
Users organize these pins into {\em boards}, which contain collections of similar pins. 
Altogether, Pinterest is the world's largest user-curated graph of images, with over $2$ billion unique pins collected into over $1$ billion boards.

Through extensive offline metrics, controlled user studies, and A/B tests, we show that our approach achieves state-of-the-art performance compared to other scalable deep content-based recommendation algorithms, in both an item-item recommendation task (\ie, related-pin recommendation), as well as a ``homefeed'' recommendation task. 
In offline ranking metrics we improve over the best performing baseline by more than 40\%, in head-to-head human evaluations our recommendations are preferred about 60\% of the time, and the A/B tests show $30\%$ to $100\%$ improvements in user engagement across various settings. 

To our knowledge, this is the largest-ever application of deep graph embeddings and paves the way for new generation of recommendation systems based on graph convolutional architectures.

\section{Related work}
\label{sec:related}

Our work builds upon a number of recent advancements in deep learning methods for graph-structured data. 

The notion of neural networks for graph data was first outlined in Gori et al. (2005) \cite{gori2005new} and further elaborated on in Scarselli et al. (2009) \cite{scarselli2009graph}.
However, these initial approaches to deep learning on graphs required running expensive neural ``message-passing'' algorithms to convergence and were prohibitively expensive on large graphs. 
Some limitations were addressed by Gated Graph Sequence Neural Networks \cite{li2015gated}---which employs modern recurrent neural architectures---but the approach remains computationally expensive and has mainly been used on graphs with ${<}10,000$ nodes. 

More recently, there has been a surge of methods that rely on the notion of ``graph convolutions'' or Graph Convolutional Networks (GCNs).
This approach originated with the work of Bruna et al. (2013), which developed a version of graph convolutions based on spectral graph thery \cite{bruna2013spectral}.
Following on this work, a number of authors proposed improvements, extensions, and approximations of these spectral convolutions \cite{berg2017graph,bronstein2017geometric,dai2016discriminative,defferrard2016convolutional,duvenaud2015convolutional,hamilton2017inductive,kipf2016semi,monti2017geometric,Zitnik2018},
leading to new state-of-the-art results on benchmarks such as node classification, link prediction, as well as recommender system tasks (\eg, the MovieLens benchmark \cite{monti2017geometric}). 
These approaches have consistently outperformed techniques based upon matrix factorization or random walks (\eg, node2vec \cite{grover2016node2vec} and DeepWalk \cite{perozzi2014deepwalk}), and their success has led to a surge of interest in applying GCN-based methods to applications ranging from recommender systems \cite{monti2017geometric} to drug design \cite{kearnes2016molecular,Zitnik2018}.
Hamilton et al. (2017b) \cite{hamilton2017representation} and Bronstein et al. (2017) \cite{bronstein2017geometric}
provide comprehensive surveys of recent advancements. 

However, despite the successes of GCN algorithms, no previous works have managed to apply them to production-scale data with billions of nodes and edges---a limitation that is primarily due to the fact that  traditional GCN methods require operating on the entire graph Laplacian during training. 
Here we fill this gap and show that GCNs can be scaled to operate in a production-scale recommender system setting involving billions of nodes/items. 
Our work also demonstrates the substantial impact that GCNs have on recommendation performance in a real-world environment.

In terms of algorithm design, our work is most closely related to Hamilton et al. (2017a)'s GraphSAGE algorithm \cite{hamilton2017inductive} and the closely related follow-up work of Chen et al. (2018) \cite{chen2018fastgcn}.
GraphSAGE is an inductive variant of GCNs that we modify to avoid operating on the entire graph Laplacian. 
We fundamentally improve upon GraphSAGE by removing the limitation that the whole graph be stored in GPU memory, using low-latency random walks to sample graph neighborhoods in a producer-consumer architecture.
We also introduce a number of new training techniques to improve performance and a MapReduce inference pipeline to scale up to graphs with billions of nodes. 

Lastly, also note that graph embedding methods like node2vec \cite{grover2016node2vec} and DeepWalk \cite{perozzi2014deepwalk} cannot be applied here. First, these are unsupervised methods. Second, they cannot include node feature information. Third, they directly learn embeddings of nodes and thus the number of model parameters is linear with the size of the graph, which is prohibitive for our setting.

\section{Method}
\label{sec:method}

\cut{
Pinterest is a website for social curation where users can interact with {\em pins}, which are visual bookmarks to online content (\textit{e.g.}, recipes they want to cook, clothes they want to purchase).
Users organize these pins into {\em boards}, which contain collections of similar pins. 
Altogether, Pinterest possesses one of the world's largest user-curated datasets of images, with over $2$ billion unique pins in the social graph. 
}

\cut{
Our primary goal is to leverage \name to develop a highly scalable representation learning system
at Pinterest to power recommendation tasks.
For example, in item-item recommendation, we want to recommend items that are closest in embedding space to
a given query pin; in user homefeed recommendation, we aim to recommend items that are closest in
embedding space to one of the pins that the user has shown interest to.
}

In this section, we describe the technical details of the \name\ architecture and training, as well as a MapReduce pipeline to efficiently generate embeddings using a trained \name\ model. 

The key computational workhorse of our approach is the notion of localized graph convolutions.\footnote{Following a number of recent works (\eg, \cite{duvenaud2015convolutional,kearnes2016molecular}) we use the term ``convolutional'' to refer to a module that aggregates information from a local graph region and to denote the fact that parameters are shared between spatially distinct applications of this module; however, the architecture we employ does not directly approximate a spectral graph convolution (though they are intimately related) \cite{bronstein2017geometric}.}
To generate the embedding for a node (\ie, an item), we apply multiple convolutional modules that aggregate feature information (\eg, visual, textual features) from the node's local graph neighborhood (Figure~\ref{fig:model}).  
Each module learns how to aggregate information from a small graph neighborhood, and by stacking multiple such modules, our approach can gain information about the local network topology.
Importantly, parameters of these localized convolutional modules are shared across all nodes, making
the parameter complexity of our approach independent of the input graph size.

\cut{
In order to scale to graphs with billions of nodes, we leverage the {\em inductive} capability of
\name, which allows this approach to generate embeddings for unseen nodes after training. 
First, we train \name using a multi-tower GPU setup on a subset of the data (${\approx}20\%$ of all nodes). Then we use \cut{the trained weights and} an efficient MapReduce inference scheme to generate embeddings for all nodes. 
}

\newcommand{\Ss}{\mathcal{I}}
\newcommand{\Cc}{\mathcal{C}}
\newcommand{\Vv}{\mathcal{V}}
\subsection{Problem Setup}\label{sec:setup}
Pinterest is a content discovery application where users interact with {\em pins}, which are visual bookmarks to online content (\textit{e.g.}, recipes they want to cook, or clothes they want to purchase).
Users organize these pins into {\em boards}, which contain collections of pins that the user deems to be thematically related.
Altogether, the Pinterest graph contains 2 billion pins, 1 billion boards, and over 18 billion edges (\ie, memberships of pins to their corresponding boards). 

Our task is to generate high-quality embeddings or representations of pins that can be used for recommendation (\eg, via nearest-neighbor lookup for related pin recommendation, or for use in a downstream re-ranking system). 
In order to learn these embeddings, we model the Pinterest  environment as a bipartite graph consisting of nodes in two disjoint sets, $\Ss$ (containing pins) and $\Cc$ (containing boards). 
Note, however, that our approach is also naturally generalizable, with $\Ss$ being viewed as a set of items and $\Cc$ as a set of user-defined contexts or collections. 

In addition to the graph structure, we also assume that the pins/items $u \in \Ss$ are associated with real-valued attributes, $x_u \in \R^d$.
In general, these attributes may specify metadata or content information about an item, and in the case of Pinterest, we have that pins are associated with both rich text and image features. 
Our goal is to leverage both these input attributes as well as the structure of the bipartite graph to generate high-quality embeddings. These embeddings are then used for recommender system candidate generation via nearest neighbor lookup (\ie, given a pin, find related pins) or as features in machine learning systems for ranking the candidates.

For notational convenience and generality, when we describe the \name~ algorithm, we simply refer to the node set of the full graph with  $\Vv = \Ss \cup \Cc$ and do not explicitly distinguish between pin and board nodes (unless strictly necessary), using the more general term ``node'' whenever possible.

\cut{
Previous \framework algorithms define neighborhood of a node $u$ as a set of nodes having an edge with $u$.
Here we propose a novel random-walk-based weighted neighborhood definition and demonstrate its use in localized graph
convolutions in Section \ref{sec:arch}. 
This technique results in significant performance improvement over the vanilla \framework
algorithms.
}


\subsection{Model Architecture}
\label{sec:arch}

We use localized convolutional modules to generate embeddings for nodes. We start with input node features and then learn neural networks that transform and aggregate features over the graph to compute the node embeddings (Figure~\ref{fig:model}).

\xhdr{Forward propagation algorithm}
We consider the task of generating an embedding, $\mb{z}_u$ for a node $u$, which depends on the node's input features and the graph structure around this node.  

\begin{algorithm}[h!]
\caption{\textsc{convolve}}
\label{alg:convolve}
\SetKwInOut{Input}{Input}\SetKwInOut{Output}{Output}
\Input{ Current embedding $\mb{z}_u$ for node $u$; set of neighbor embeddings $\{\mb{z}_v | v \in \mathcal{N}(u)\}$, set of neighbor weights $\bm{\alpha}$;
symmetric vector function $\gamma(\cdot)$}
\Output{ New embedding $\mb{z}^{\textsc{new}}_u$ for node $u$}
 \BlankLine
  $	\mb{n}_{u} \leftarrow \gamma\left(\left\{\textrm{ReLU}\left(\mb{Q} \mb{h}_v + \mb{q} \right)
      \mid v \in \mathcal{N}(u) \right\}, \bm{\alpha} \right)$\;
    	$\mb{z}^{\textsc{new}}_u\leftarrow \textrm{ReLU}\left(\mb{W}\cdot\textsc{concat}(\mb{z}_u, \mb{n}_u) + \mb{w}\right)$\;
    	$\mb{z}^{\textsc{new}}_u\leftarrow \mb{z}^{\textsc{new}}_u / \| \mb{z}^{\textsc{new}}_u\|_2$
\end{algorithm}

The core of our \name~ algorithm is a localized convolution operation, where we learn how to aggregate information from $u$'s neighborhood (Figure \ref{fig:model}).
This procedure is detailed in Algorithm \ref{alg:convolve} \textsc{convolve}.
The basic idea is that we transform the representations $\mb{z}_v, \forall v \in \mathcal{N}(u)$ of $u$'s neighbors 
through a dense neural network and then apply a aggregator/pooling fuction (\textit{e.g.}, a
element-wise mean or weighted sum, denoted as $\gamma$) on the resulting set of vectors (Line 1). 
This aggregation step provides a vector representation, $\mb{n}_u$, of $u$'s local neighborhood, $\mathcal{N}(u)$.
We then concatenate the aggregated neighborhood vector $\mb{n}_u$ with $u$'s current representation $\mb{h}_u$ and transform the concatenated vector through another dense neural network layer (Line 2).
Empirically we observe significant performance gains when using concatenation operation instead of the average operation as in \cite{kipf2016semi}.
Additionally, the normalization in Line 3 makes training more stable, and it is more efficient to perform approximate nearest neighbor search for normalized embeddings (Section \ref{sec:nearest_neighbor_lookups}).
The output of the algorithm is a representation of $u$ that incorporates both information about itself and its local graph neighborhood. 

\xhdr{Importance-based neighborhoods}
An important innovation in our approach is how we define node neighborhoods $\mathcal{N}(u)$, \ie, how we select the set of neighbors to convolve over in Algorithm \ref{alg:convolve}.
Whereas previous GCN approaches simply examine $k$-hop graph neighborhoods, in \name we define importance-based neighborhoods, where the neighborhood of a node $u$ is defined as the $T$ nodes that exert the most influence on node $u$.
Concretely, we simulate random walks starting from node $u$ and compute the $L_1$-normalized visit count of nodes visited by
the random walk~\cite{pixie2018}.\footnote{In the limit of infinite simulations, the normalized counts approximate
  the Personalized PageRank scores with respect to $u$.}
The neighborhood of $u$ is then defined as the top $T$ nodes with the highest normalized visit counts with respect to node $u$.

The advantages of this importance-based neighborhood definition are two-fold.
First, selecting a fixed number of nodes to aggregate from allows us to control the memory footprint of the algorithm during training \cite{hamilton2017inductive}.
Second, it allows Algorithm \ref{alg:convolve} to take into account the importance of neighbors when aggregating the vector representations of neighbors.
In particular, we implement $\gamma$ in Algorithm \ref{alg:convolve} as a weighted-mean, with weights defined according to the $L_1$ normalized visit counts.
We refer to this new approach as {\em importance pooling}. 

\xhdr{Stacking convolutions}
Each time we apply the \textsc{convolve} operation (Algorithm \ref{alg:convolve}) we get a new representation for a node, and
we can stack multiple such convolutions on top of each other in order to gain more information about the local graph structure around node $u$. 
In particular, we use multiple layers of convolutions, where the inputs to the convolutions at layer $k$ depend on the representations output from layer $k-1$ (Figure \ref{fig:model}) and where the initial (\ie, ``layer 0'') representations are equal to the input node features. 
Note that the model parameters in Algorithm \ref{alg:convolve} ($\mb{Q}$, $\mb{q}$, $\mb{W}$, and $\mb{w}$) are shared across the nodes but differ between layers. 

Algorithm \ref{alg:minibatch} details how stacked convolutions generate embeddings for a minibatch set of nodes, $\mathcal{M}$. 
We first compute the neighborhoods of each node and then apply $K$ convolutional iterations to generate the layer-$K$ representations of the target nodes.
The output of the final convolutional layer is then fed through a fully-connected neural network to generate the final output embeddings $\mb{z}_u, \forall u \in \mathcal{M}$. 

The full set of parameters of our model which we then learn is: the weight and bias parameters for
each convolutional layer ($\mb{Q}^{(k)}, \mb{q}^{(k)}, \mb{W}^{(k)}, \mb{w}^{(k)}, \forall k \in
\{1,...,K\}$) as well as the parameters of the final dense neural network layer, $\mb{G}_1$, $\mb{G}_2$, and $\mb{g}$. 
The output dimension of Line 1 in Algorithm \ref{alg:convolve} (\ie, the column-space dimension of $\mb{Q}$) is set to be $m$ at all layers.
For simplicity, we set the output dimension of all convolutional layers (\ie, the output at Line 3 of Algorithm \ref{alg:convolve}) to be equal, and we denote this size parameter by $d$. 
The final output dimension of the model (after applying line 18 of Algorithm \ref{alg:minibatch}) is also set to be $d$. 

\begin{algorithm}[t!]
    \caption{\textsc{minibatch}}
    \label{alg:minibatch}
	\SetKwInOut{Input}{Input}\SetKwInOut{Output}{Output}
    \Input{Set of nodes $\mathcal{M} \subset \mathcal{V}$; depth parameter $K$; neighborhood function $\mathcal{N} : \mathcal{V} \rightarrow 2^\mathcal{V}$}
    \Output{Embeddings $\mb{z}_u, \forall u \in \mathcal{M}$}
    \BlankLine
    \tcc{Sampling neighborhoods of minibatch nodes.}
     $\mathcal{S}^{(K)} \leftarrow \mathcal{M}$\;
     \For{$k=K,...,1$}{
        $\mathcal{S}^{(k-1)} \leftarrow \mathcal{S}^{(k)}$\;
     	 \For{$u \in S^{(k)}$}{
     	 	$\mathcal{S}^{(k-1)} \leftarrow \mathcal{S}^{(k-1)} \cup \mathcal{N}(u)$\;
     	 }
    }
    \tcc{Generating embeddings}
   $ \mb{h}_u^{(0)} \leftarrow \mb{x}_u, \forall u \in \mathcal{S}^{(0)}$\;
    \For{$k=1,...,K$}{
    	\For{$u \in \mathcal{S}^{(k)}$}{
    	    $\mathcal{H} \leftarrow \left\{\mb{h}_v^{(k-1)}, \forall v \in \mathcal{N}(u)\right\}$\;
    	    $ \mb{h}_u^{(k)} \leftarrow \textsc{convolve}^{(k)}\left(\mb{h}_u^{(k-1)}, \mathcal{H}\right)$
    	}
    }
    \For{$u \in \mathcal{M}$}{
         $\mb{z}_u \leftarrow \mb{G}_2\cdot\textrm{ReLU}\left(\mb{G}_1\mb{h}^{(K)}_u + \mb{g}\right)$
     }      
\end{algorithm}


\cut{
\begin{itemize}
\item We use the GraphSAGE framework (with bias), with 2 layers, and mean-pooling aggregators. \will{We should fully describe this, and I think we should just talk about the minibatch version, even though it is messier.}
\item Input feature dimension: 4096 + 256 + 1 (visual embedding, annotation embedding and log degree of node)
\item Neighborhood features and self features are concatenated
\item Hidden dimension is 1024; pooling dimension is 2048; output dimension is 1024.
\end{itemize}
}

\subsection{Model Training}
\label{sec:training}

We train \name\ in a supervised fashion using a max-margin ranking loss. 
In this setup, we assume that we have access to a set of labeled pairs of items $\mathcal{L}$, where the pairs in the set, $(q,i) \in \mathcal{L}$, are assumed to be related---\ie, we assume that if $(q,i) \in \mathcal{L}$ then item $i$ is a good recommendation candidate for query item $q$. 
The goal of the training phase is to optimize the \name\ parameters so that the output embeddings of pairs $(q,i) \in \mathcal{L}$ in the labeled set are close together.

We first describe our margin-based loss function in detail.
Following this, we give an overview of several techniques we developed that lead to the computation efficiency and fast convergence rate of \name, allowing us to train on billion node graphs and billions training examples.
And finally, we describe our curriculum-training scheme, which improves the overall quality of the recommendations. 

\xhdr{Loss function}
In order to train the parameters of the model, we use a max-margin-based loss function.
The basic idea is that we want to maximize the inner product of positive examples, \ie, the embedding of the query item and the corresponding related item.
At the same time we want to ensure that the inner product of negative examples---\ie, the inner product between the embedding of the query item and an unrelated item---is smaller than that of the positive sample by some pre-defined margin. 
The loss function for a single pair of node embeddings $(\mb{z}_q, \mb{z}_i) :  (q,i) \in \mathcal{L}$ is thus
\begin{equation}
    J_{\mathcal{G}}(\mb{z}_q \mb{z}_i) = \mathbb{E}_{n_k \sim P_n(q)} \max\{0, \mb{z}_q \cdot \mb{z}_{n_k} - \mb{z}_q\cdot \mb{z}_i + \Delta\},
    \label{eq:loss}
\end{equation}
where $P_n(q)$ denotes the distribution of negative examples for item $q$, and $\Delta$ denotes the margin hyper-parameter. 
We shall explain the sampling of negative samples below.

\xhdr{Multi-GPU training with large minibatches}
To make full use of multiple GPUs on a single machine for training, we run the forward and backward propagation in a multi-tower fashion.
With multiple GPUs, we first divide each minibatch (Figure~\ref{fig:model} bottom) into equal-sized portions. 
Each GPU takes one portion of the minibatch and performs the computations using the same set of parameters.
After backward propagation, the gradients for each parameter across all GPUs are aggregated together, and a single step of synchronous SGD is performed.
Due to the need to train on extremely large number of examples (on the scale of billions), we run our system with large batch sizes, ranging from $512$ to $4096$.

We use techniques similar to those proposed by Goyal {\em et al.} \cite{goyal2017accurate} to ensure fast convergence and maintain training and generalization accuracy when dealing with large batch sizes.
We use a gradual warmup procedure that increases learning rate from small to a peak value in the
first epoch according to the linear scaling rule. Afterwards the learning rate is decreased exponentially.

\xhdr{Producer-consumer minibatch construction}
During training, the adjacency list and the feature matrix for billions of nodes are placed in CPU memory due to their large size. 
However, during the \textsc{convolve} step of \name, each GPU process needs access to the 
neighborhood and feature information of nodes in the neighborhood. Accessing the data in CPU memory from GPU is not efficient. To solve this problem, we use a \emph{re-indexing} technique to create a sub-graph $G'=(V', E')$ containing nodes and their neighborhood, which will be involved in the computation of the current minibatch. 
\cut{The node set is $V' = \bigcup_{u \in \mathcal{M}} \mathcal{N}(u)$---the union of neighborhood for all nodes in the current minibatch $\mathcal{M}$---and $E' = \{(u, v) : u \in V', v \in V', (u, v) \in E\}$. }
A small feature matrix containing only node features relevant to computation of the current minibatch is also extracted such that the order is consistent with the index of nodes in $G'$.
The adjacency list of $G'$ and the small feature matrix are fed into GPUs at the start of each minibatch iteration, so that no communication between the GPU and CPU is needed during the \textsc{convolve} step, greatly improving GPU utilization.

The training procedure has alternating usage of CPUs and GPUs.  
The model computations are in GPUs, whereas extracting features, re-indexing, and negative sampling are computed on CPUs. 
In addition to parallelizing GPU computation with multi-tower training, and CPU computation using OpenMP \cite{openmp15}, 
we design a producer-consumer pattern to run GPU computation at the current iteration and CPU computation at the next iteration in parallel. This further reduces the training time by almost a half.

\xhdr{Sampling negative items}
Negative sampling is used in our loss function (Equation \ref{eq:loss}) as an approximation of the normalization factor of edge likelihood \cite{mikolov2013distributed}.
To improve efficiency when training with large batch sizes, we sample a set of 500 negative items to
be shared by all training examples in each minibatch.
This drastically saves the number of embeddings that need to be computed during each training step, compared to running negative sampling for each node independently.
Empirically, we do not observe a difference between the performance of the two sampling schemes.

In the simplest case, we could just uniformly sample negative examples from the entire set of items. 
However, ensuring that the inner product of the positive example (pair of items $(q,i)$) is larger than that of the $q$ and each of the 500 negative items is too ``easy'' and does not provide fine enough ``resolution'' for the system to learn.
In particular, our recommendation algorithm should be capable of finding 1,000 most relevant items to $q$ among the catalog of over 2 billion items. In other words, our model should be able to distinguish/identify 1 item out of 2 million items.
But with 500 random negative items, the model's resolution is only 1 out of 500. 
Thus, if we sample 500 random negative items out of 2 billion items, the chance of any of these items being even slightly related to the query item is small. 
Therefore, with large probability the learning will not make good parameter updates and will not be able to differentiate slightly related items from the very related ones.

\begin{figure}[t]
  \centering
  \includegraphics[width=0.45\textwidth]{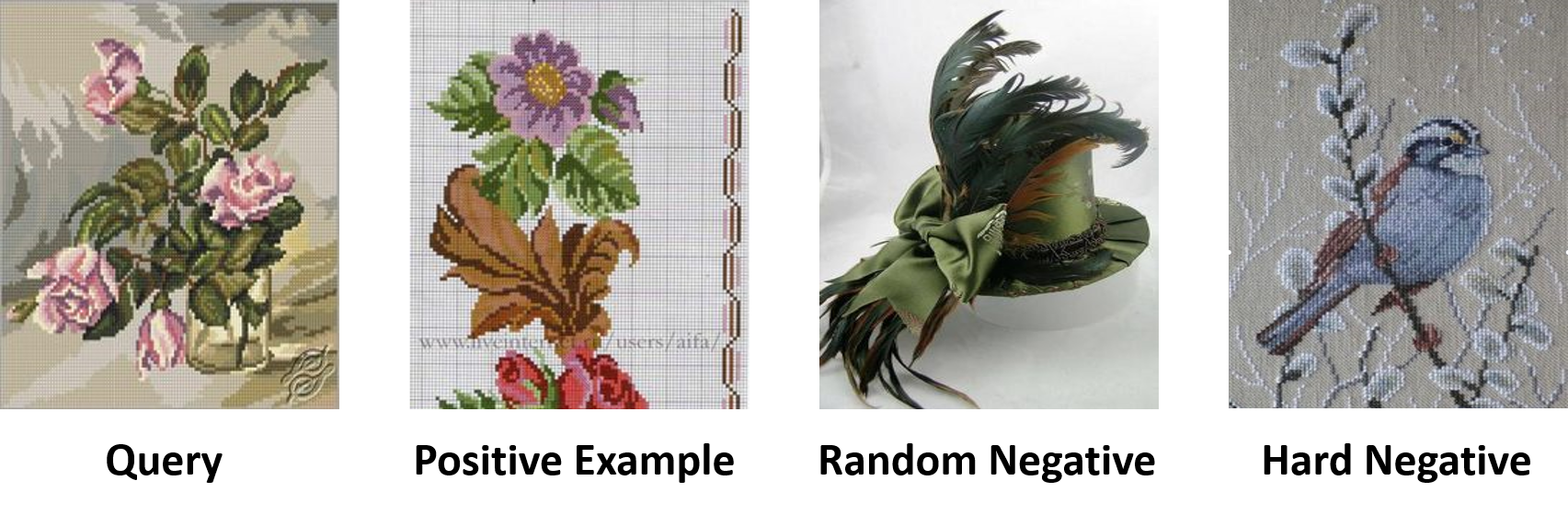}
  \caption{Random negative examples and hard negative examples. Notice that the hard negative example is significantly more similar to the query, than the random negative example, though not as similar as the positive example. }
  \label{fig:hard_neg_examples}
\end{figure}

To solve the above problem, for each positive training example (\ie, item pair $(q, i)$), we add ``hard'' negative examples, \ie, items that are somewhat related to the query item $q$, but not as related as the positive item $i$.
We call these ``hard negative items''. They are generated by ranking items in a graph according to their Personalized PageRank scores with respect to query item $q$~\cite{pixie2018}.
Items ranked at 2000-5000 are randomly sampled as hard negative items. As illustrated in Figure \ref{fig:hard_neg_examples}, the hard negative examples are more similar to the query than random negative examples, and are thus challenging for the model to rank, forcing the model to learn to distinguish items 
at a finer granularity.

Using hard negative items throughout the training procedure doubles the number of epochs needed for the training to converge.
To help with convergence, we develop a curriculum training scheme~\cite{bengio2009curriculum}. In
the first epoch of training, no hard negative items are used, so that the algorithm quickly finds an
area in the parameter space where the loss is relatively small. We then add hard negative items in
subsequent epochs, focusing the model to learn how to distinguish highly related pins from only slightly related ones. At epoch $n$ of the training, we add $n-1$ hard negative items to the set of negative items for each item.

\subsection{Node Embeddings via MapReduce}
\label{sec:inference}
After the model is trained, it is still challenging to directly apply the trained model to generate
embeddings for all items, including those that were not seen during training. 
Naively computing embeddings for nodes using Algorithm~\ref{alg:minibatch}
leads to repeated computations caused by the overlap between $K$-hop neighborhoods of nodes.
As illustrated in Figure \ref{fig:model}, many nodes are repeatedly computed at multiple layers when generating the embeddings for different target nodes. 
To ensure efficient inference, we develop a MapReduce approach that runs model inference without repeated computations.

\cut{
Algorithm \ref{alg:mapreduce} outlines the basic MapReduce algorithm for \name\ (excluding details on data flow, etc.).
\begin{algorithm}
\caption{\textsc{MapReduce}}
\label{alg:mapreduce}
\SetKwInOut{Input}{Input}\SetKwInOut{Output}{Output}
    \Input{Vertex set $\mathcal{V}$; depth parameter $K$}
    \Output{Embeddings $\mb{z}_u, \forall u \in \mathcal{V}$}
  \SetKwFunction{algo}{MapReduce}\SetKwFunction{map}{map}\SetKwFunction{reduce}{reduce$^{(k)}$}
  \SetKwProg{myalg}{Algorithm}{}{}
  \myalg{\algo{$\mathcal{V}$, $K$}}{
  $\mathcal{H}^{(0)} \leftarrow \{(u, \mb{x}_u), \forall u \in \mathcal{V}\}$\;
  \For{$k=1,...,K$}{
  	 $\mathcal{H}^{(k)} \leftarrow \texttt{reduce}\left(\texttt{map}\left(\mathcal{H}^{(k-1)}\right)\right)$
  }
   \For{$\mb{h}^{(K)}_u \in \mathcal{H}^{(K)}$}{
         $\mb{z}_u \leftarrow \mb{G}_2\cdot\textrm{ReLU}\left(\mb{G}_1\mb{h}^{(K)}_u + \mb{g}\right)$
     }     
  }
  \setcounter{AlgoLine}{0}
  \SetKwProg{myproc}{Procedure}{:}{}
  \myproc{\map{$u, \mb{h}_u$}}{
  	\For{$v \in \mathcal{N}(u) \cup \{u\}$} {
  		\textbf{output} $\textrm{key}=v$, $\textrm{value}=(u, \mb{h}_u)$\;
 	 }
  }
  \setcounter{AlgoLine}{0}
  \SetKwProg{myproc}{Procedure}{:}{}
  \myproc{\reduce{$u, \left\{(v, \mb{h}_v) \: \forall v \in \mathcal{N}(u)\cup\{u\}\right\}$}}{
     $ \mb{h}_u^{\textrm{new}} = \textsc{convolve}^{(k)}(\mb{h}_u, \left\{ \mb{h}_v \: \forall v \in \mathcal{N}(u)\right\})$\;
     \textbf{output} $\textrm{key}=u$, $\textrm{value}=\mb{h}^{\textrm{new}}_u$\;
  	}
\ref{alg:mapreduce}
\end{algorithm}
} 

\begin{figure*}[t]
    \centering
    \includegraphics[width=0.95\textwidth]{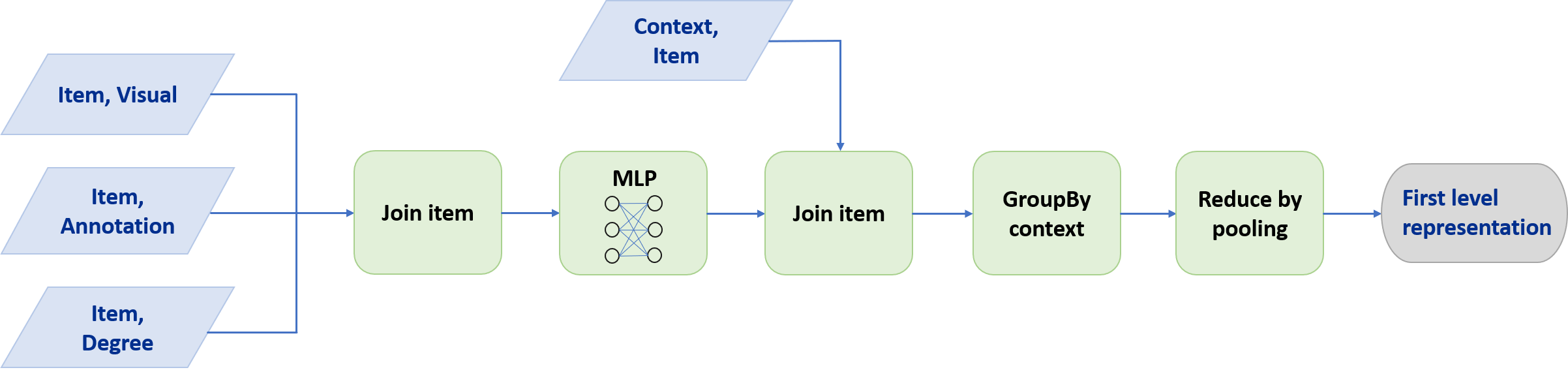}
    \caption{Node embedding data flow to compute the first layer representation using MapReduce. The second layer computation follows the same pipeline, except that the inputs are first layer representations, rather than raw item features.}
    \label{fig:inference_dataflow}
\end{figure*}

We observe that inference of node embeddings very nicely lends itself to MapReduce computational model.  
Figure \ref{fig:inference_dataflow} details the data flow on the bipartite pin-to-board Pinterest graph, where we assume the input (\ie, ``layer-0'') nodes are pins/items (and the layer-1 nodes are boards/contexts). 
The MapReduce pipeline has two key parts:
\begin{enumerate}
    \item One MapReduce job is used to project all pins to a low-dimensional latent space, where the aggregation operation will be performed (Algorithm~\ref{alg:convolve}, Line 1). 
    \item Another MapReduce job is then used to join the resulting pin representations with the ids of the boards they occur in, and the board embedding is computed by pooling the features of its (sampled) neighbors.
\end{enumerate}
Note that our approach avoids redundant computations and that the latent vector for each node is computed only once. 
After the embeddings of the boards are obtained, we use two more MapReduce jobs to compute the second-layer embeddings of pins, in a similar fashion as above, and this process can be iterated as necessary (up to $K$ convolutional layers).\footnote{Note that since we assume that only pins (and not boards) have features, we must use an even number of convolutional layers.} 
%

\subsection{Efficient nearest-neighbor lookups}
\label{sec:nearest_neighbor_lookups}

The embeddings generated by \name\ can be used for a wide range of downstream recommendation tasks, and in many settings we can directly use these embeddings to make recommendations by performing nearest-neighbor lookups in the learned embedding space. 
That is, given a query item $q$, the we can recommend items whose embeddings are the $K$-nearest neighbors of the query item's embedding.
Approximate KNN can be obtained efficiently via locality sensitive hashing \cite{andoni2006near}.  
After the hash function is computed, retrieval of items can be implemented with a two-level retrieval process based on the Weak AND operator \cite{broder2003efficient}.
Given that the \name model is trained offline and all node embeddings are computed via MapReduce and saved in a database, the efficient nearest-neighbor lookup operation enables the system to serve recommendations in an online fashion, 


\section{Experiments}
\label{sec:experiments}

To demonstrate the efficiency of \name \ and the quality of the embeddings it generates, we conduct a comprehensive suite of experiments on the entire Pinterest object graph, including offline experiments, production A/B tests as well as user studies.

\subsection{Experimental Setup}
\label{sec:setup}

\cut{
Users of Pinterest curate visual objects called pins into collections called boards. So, each node in this graph either represents a pin (\ie, item set $\Ss$), or a board (\ie, context set $\Cc)$. 
An edge connecting a board node to a pin node indicates that a user has placed that pin on the given board. 
}

We evaluate the embeddings generated by \name\ in two tasks: recommending related pins and recommending pins in a user's home/news feed.
To recommend related pins, we select the $K$ nearest neighbors to the query pin in the embedding space. 
We evaluate performance on this related-pin recommendation task using both offline ranking measures as well as a controlled user study.
For the homefeed recommendation task, we select the pins that are closest in the embedding space to one of the most recently pinned items by the user. 
We evaluate performance of a fully-deployed production system on this task using A/B tests to measure the overall impact on user engagement.

\xhdr{Training details and data preparation}
We define the set, $\mathcal{L}$, of positive training examples (Equation \eqref{eq:loss}) using historical user engagement data. 
In particular, we use historical user engagement data to identify pairs of pins $(q,i)$, where a user interacted with pin $i$ immediately after she interacted with pin $q$. We use all other pins as negative items (and sample them as described in Section~\ref{sec:training}). Overall, we use $1.2$ billion pairs of positive training examples (in addition to $500$ negative examples per batch and $6$ hard negative examples per pin). Thus in total we use $7.5$ billion training examples. 

Since \name\ can efficiently generate embeddings for unseen data, we only train on a subset of the Pinterest graph and then generate embeddings for the entire graph using the MapReduce pipeline described in Section \ref{sec:inference}. 
In particular, for training we use a randomly sampled subgraph of the entire graph, containing $20\%$ of all boards (and all the pins touched by those boards) and $70\%$ of the labeled examples.  
During hyperparameter tuning, a remaining $10\%$ of the labeled examples are used. And, when testing, we run inference on the entire graph to compute embeddings for all $2$ billion pins, and the remaining $20\%$ of the labeled examples are used to test the recommendation performance of our \name in the offline evaluations.
Note that training on a subset of the full graph drastically decreased training time, with a negligible impact on final performance. 
In total, the full datasets for training and evaluation are approximately $18$TB in size with the full output embeddings being $4$TB. 

\xhdr{Features used for learning}
Each pin at Pinterest is associated with an image and a set of textual annotations (title, description). 
To generate feature representation $\mb{x}_q$ for each pin $q$, we concatenate visual embeddings (4,096 dimensions), textual annotation embeddings (256 dimensions), and the log degree of the node/pin in the graph. 
The visual embeddings are the 6-th fully connected layer of a classification network using the VGG-16 architecture \cite{simonyan2014very}. 
Textual annotation embeddings are trained using a Word2Vec-based model \cite{mikolov2013distributed}, where the context of an annotation consists of other annotations that are associated with each pin.

\xhdr{Baselines for comparison}
We evaluate the performance of \name against the following state-of-the-art content-based, graph-based and deep learning baselines that generate embeddings of pins:
\begin{enumerate}[topsep=3pt, leftmargin=15pt]
	\item Visual embeddings (\textbf{Visual}): Uses nearest neighbors of deep visual embeddings for
      recommendations. The visual features are described above.
	\item Annotation embeddings (\textbf{Annotation}): Recommends based on nearest neighbors in terms of annotation embeddings. The annotation embeddings are described above.
	\item Combined embeddings (\textbf{Combined}): Recommends based on  concatenating visual and annotation embeddings, and using a 2-layer multi-layer perceptron to compute embeddings that capture both visual and annotation features.
	\item Graph-based method (\textbf{Pixie}): This random-walk-based method \cite{pixie2018} uses biased random walks to
    generate ranking scores by simulating random walks starting at query pin $q$. Items with top $K$
    scores are retrieved as recommendations. While this approach does not generate pin embeddings, it is currently the state-of-the-art at Pinterest for certain recommendation tasks \cite{pixie2018} and thus an informative baseline. 
\end{enumerate}
The visual and annotation embeddings are state-of-the-art deep learning content-based systems currently deployed at Pinterest to generate representations of pins. 
Note that we do not compare against other deep learning baselines from the literature simply due to the scale of our problem. 
We also do not consider non-deep learning approaches for generating item/content embeddings, since other works have already proven state-of-the-art performance of deep learning approaches for generating such embeddings \cite{covington2016deep,monti2017geometric,van2013deep}.
\cut{
We also considered comparing against other state-of-the-art deep recommender system baselines that
incorporate graph information,
including \cite{li2017collaborative,wu2016collaborative}.
However, due to the non-inductive nature of these algorithms, training 3 billion dense embeddings
involves too many parameters, resulting in prohibitively long training time for
\cite{li2017collaborative}.
The denoising autoencoder proposed in \cite{wu2016collaborative} also cannot meet our need since its
input dimension is proportional to the number of items.
}

We also conduct ablation studies and consider several variants of \name~ when evaluating performance:
\begin{itemize}[topsep=3pt, leftmargin=15pt]
	\item {\bf max-pooling }uses the element-wise max as a symmetric aggregation function (i.e., $\gamma=\max$) without hard negative samples; 
	\item {\bf mean-pooling} uses the element-wise mean as a symmetric aggregation function (i.e., $\gamma=\textrm{mean}$);
	\item {\bf mean-pooling-xent} is the same as mean-pooling but uses the cross-entropy loss introduced in \cite{hamilton2017inductive}.
	\item {\bf mean-pooling-hard} is the same as mean-pooling, except that it incorporates hard negative samples as detailed in Section \ref{sec:training}.
	\item {\bf \name} uses all optimizations presented in this paper, including the use of importance pooling in the convolution step.
\end{itemize}
The max-pooling and cross-entropy settings are extensions of the best-performing GCN model from Hamilton et al. \cite{hamilton2017inductive}---other variants (\eg, based on Kipf et al. \cite{kipf2016semi}) performed significantly worse in development tests and are omitted for brevity.\footnote{Note that the recent GCN-based recommender systems of Monti et al. \cite{monti2017geometric} and Berg et al. \cite{berg2017graph} are not directly comparable because they cannot scale to the Pinterest size data.}
For all the above variants, we used $K=2$, hidden dimension size $m = 2048$, and set the embedding dimension $d$ to be $1024$.

\xhdr{Computation resources}
\label{sec:machinery}
Training of \name is implemented in TensorFlow \cite{abadi2016tensorflow} and run on a single machine with 32 cores and 16 Tesla K80 GPUs. To ensure fast fetching of item's visual and annotation features, we store them in main memory, together with the graph, using Linux HugePages to increase the size of virtual memory pages from 4KB to 2MB.
The total amount of memory used in training is 500GB. Our MapReduce inference pipeline is run on a Hadoop2 cluster with 378 d2.8xlarge Amazon AWS nodes.

\subsection{Offline Evaluation}
To evaluate performance on the related pin recommendation task, we define the notion of \emph{hit-rate}.
 For each positive pair of pins $(q, i)$ in the test set, we use $q$ as a query pin and then compute its top $K$ nearest neighbors $\mathrm{NN}_q$ from a sample of $5$ million test pins. We then define the hit-rate as the fraction of queries $q$ where $i$ was ranked among the top $K$ of the test sample (\ie, where $i \in \mathrm{NN}_q$). This metric directly measures the probability that recommendations made by the algorithm contain the items related to the query pin $q$. In our experiments $K$ is set to be $500$.

We also evaluate the methods using Mean Reciprocal Rank (MRR), which takes into account of the rank of the item $j$ among recommended items for query item $q$:
\begin{equation}
	\mathrm{MRR} = \frac{1}{n} \sum_{(q, i) \in \mathcal{L}} \frac{1}{\left \lceil R_{i, q} / 100 \right \rceil}.
	\label{eq:mrr}
\end{equation}
Due to the large pool of candidates (more than 2 billion), we use a scaled version of the MRR in
Equation (\ref{eq:mrr}),  where $R_{i, q}$ is the rank of item $i$ among recommended items for query
$q$, and $n$ is the total number of labeled item pairs. The scaling factor $100$ ensures that, for
example, the difference between rank at $1,000$ and rank at $2,000$ is still noticeable, instead of being very close to $0$.

\begin{table}[t]
\begin{tabular}{c | c | c}
	\hline
	Method & Hit-rate & MRR \\
	\hline
	Visual & $17\%$ &  0.23 \\
	Annotation & $14\%$ & 0.19 \\
	Combined & $27\%$ & 0.37  \\
	\hline
	max-pooling & $39\%$ & 0.37 \\
	mean-pooling & $41\%$ & 0.51 \\
	mean-pooling-xent & $29\%$ & 0.35 \\
	mean-pooling-hard & $46\%$ & 0.56 \\
	\name & $\mb{67\%}$ & {\bf 0.59} \\
	\hline
\end{tabular}
\caption{Hit-rate and MRR for \name and content-based deep learning baselines.
Overall, \name gives 150\% improvement in hit rate and 60\% improvement in MRR over the best baseline.\fnref{note:no_pixie}
}
\label{fig:hit_rate_comparison}
\vspace{-7mm}
\end{table}

\cut{
\begin{table}[t]
\begin{tabular}{c | c | c}
	\hline
	Method & Hit-rate \@ 1000 & MRR \\
	\hline
	Visual & $0.023\%$ &  0.038 \\
	Annotation & $0.055\%$ & 0.013 \\
	\hline
	\name & $\mb{0.505\%}$ & {\bf 0.326} \\
	\hline
\end{tabular}
\caption{Hit-rate and MRR for \name and content-based deep learning baselines.
}
\label{fig:hit_rate_comparison}
\vspace{-8mm}
\end{table}
}

Table \ref{fig:hit_rate_comparison} compares the performance of the various approaches using the hit rate as well as the MRR.\footnote{\label{note:no_pixie}Note that we do not include the Pixie baseline in these offline comparisons because the Pixie algorithm runs in production and is ``generating'' labeled pairs $(q,j)$ for us---\ie, the labeled pairs are obtained from historical user engagement data in which the Pixie algorithm was used as the recommender system. Therefore, the recommended item $j$ is always in the recommendations made by the Pixie algorithm. However, we compare to the Pixie algorithm using human evaluations in Section \ref{sec:human_eval}.}
\name  with our new importance-pooling aggregation and hard negative examples achieves the best performance at 67\% hit-rate and 0.59 MRR, outperforming the top baseline by 40\%
absolute (150\% relative) in terms of the hit rate and also 22\% absolute (60\% relative) in terms of MRR. We also observe that combining visual and textual information works much better than using either one alone (60\% improvement of the combined approach over visual/annotation only).

\begin{figure}[t]
\centering
	\includegraphics[width=0.4\textwidth]{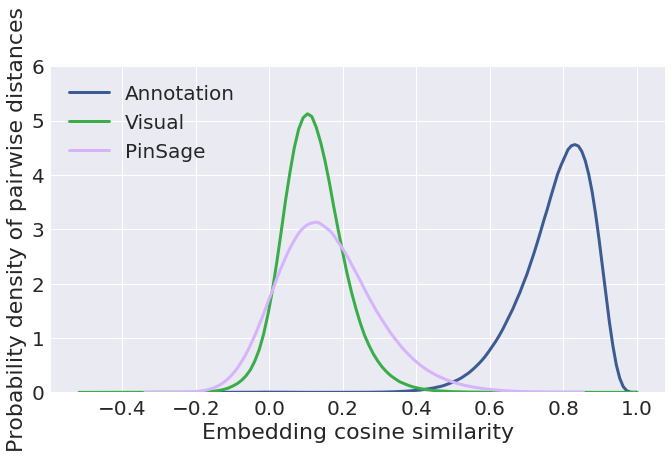}
	\caption{Probability density of pairwise cosine similarity for visual embeddings, annotation embeddings, and \name~ embeddings.}
	\label{fig:embedding_cosine_sim_dist}
	\vspace{-4mm}
\end{figure}

\xhdr{Embedding similarity distribution}
Another indication of the effectiveness of the learned embeddings is that the distances between random pairs of item embeddings are widely distributed. 
If all items are at about the same distance (\ie, the distances are tightly clustered) then the embedding space does not have enough ``resolution'' to distinguish between items of different relevance.
Figure \ref{fig:embedding_cosine_sim_dist} plots the distribution of cosine similarities between pairs of items using annotation, visual, and \name embeddings. 
This distribution of cosine similarity between random pairs of items demonstrates the effectiveness of \name, which has the most spread out distribution. In particular, the kurtosis of the cosine similarities of \name embeddings is $0.43$, compared to $2.49$ for annotation embeddings and $1.20$ for visual embeddings. 

Another important advantage of having such a wide-spread in the embeddings is that it reduces the collision probability of the subsequent LSH algorithm, thus increasing the efficiency of serving the nearest neighbor pins during
recommendation.

\subsection{User Studies}
\label{sec:human_eval}

We also investigate the effectiveness of \name by performing head-to-head comparison between different learned representations. In the user study, a user is presented with an image of
the query pin, together with two pins retrieved by two different recommendation algorithms. The user is then asked to choose which of the two candidate pins is more related to the query pin.
Users are instructed to find various correlations between the recommended items and the query item, in aspects such as visual appearance, object category and personal identity.
If both recommended items seem equally related, users have the option to choose ``equal''. If no
consensus is reached among $2/3$ of users who rate the same question, we deem the result as inconclusive.

\cut{
To assess how the relevance degrades with rank, we compare items ranked at the $10$-th, $50$-th,
$500$-th and $1,000$-th position by each method. That is, we run separate evaluations comparing rank
$10$ item of every method, and then a different evaluation comparing rank $50$ item of every method, etc.
For our model \name, we used the variant mean-pooling-hard for all
comparisons. Overall, we use 5,000 query pins and all together perform 20,000 crowdsourcing tasks/comparisons.

In this study, we compare against the previously described deep learning baselines, as well as a random-walk-based approach (Pixie) \cite{pixie2018}, which is used in various aspects of Pinterest's recommender system ecosystem and currently state-of-the-art for related-pin recommendation.
Note that we omit Pixie from the offline evaluations because it was involved in generation of the labeled data. 
} 

\begin{figure}[t]
\centering
	\begin{subfigure}[b]{0.40\textwidth}
	\includegraphics[width=1\textwidth]{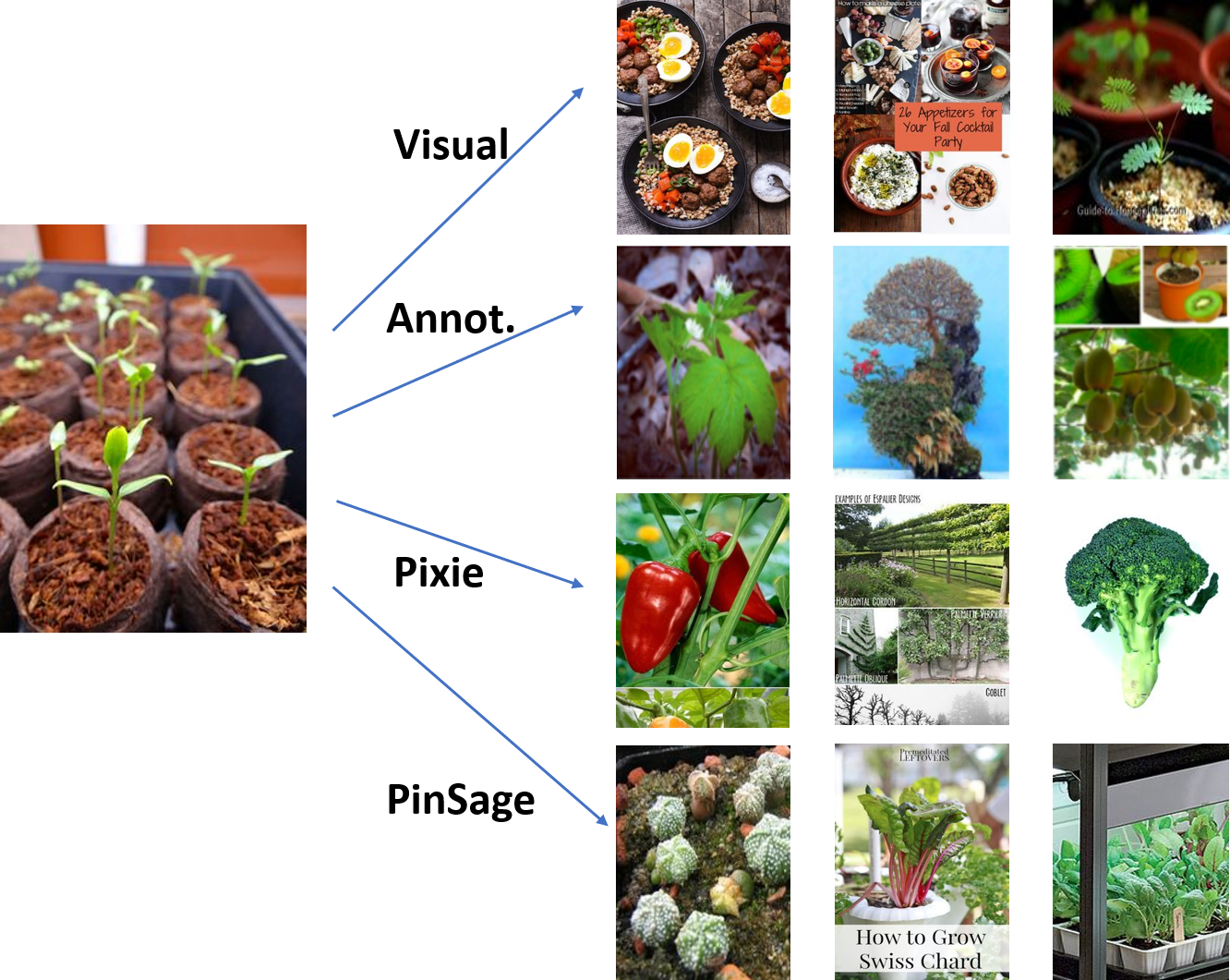}
	\end{subfigure}
	\\~\\
	\begin{subfigure}[b]{0.40\textwidth}
	\includegraphics[width=1\textwidth]{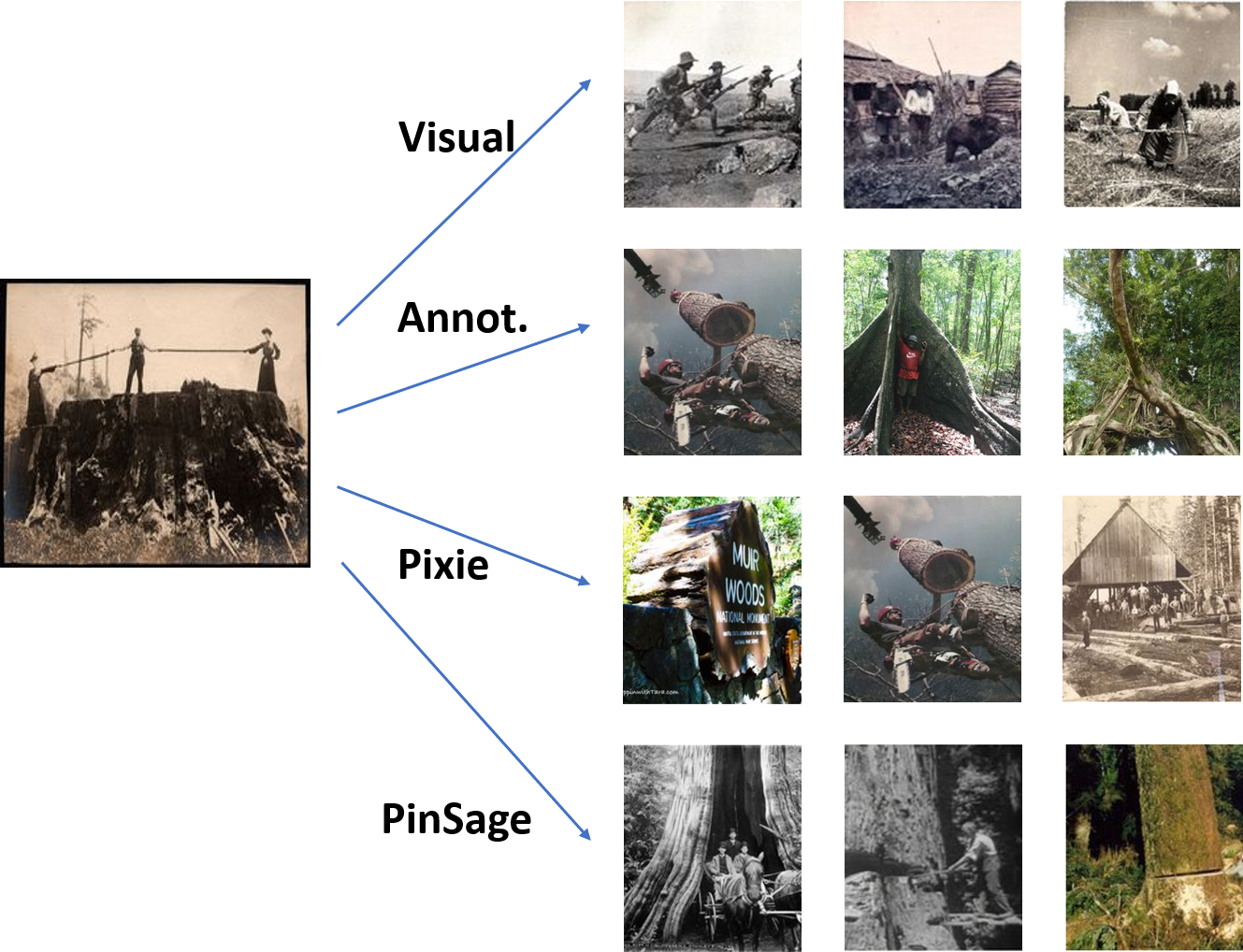}
	\end{subfigure}
	\caption{Examples of Pinterest pins recommended by different algorithms. The image to the left is the query pin. Recommended items to the right are computed using Visual embeddings, Annotation embeddings, graph-based Pixie, and \name. }
	\label{fig:rec_examples}
	\vspace{-5mm}
\end{figure}

Table \ref{fig:user_study_head2head} shows the results of the head-to-head comparison between \name~ and the 4 baselines.
Among items for which the user has an opinion of which is more related, around $60\%$ of the preferred items are recommended by \name.
Figure \ref{fig:rec_examples} gives examples of recommendations and illustrates strengths and weaknesses of the different methods. 
The image to the left represents the query item. Each row to the right corresponds to the top recommendations made by the visual embedding baseline, annotation embedding baseline, Pixie, and \name.
Although visual embeddings generally predict categories and visual similarity well, they occasionally make large mistakes in terms of image semantics. In this example, visual information confused plants with food, and tree logging with war photos, due to similar image style and appearance. 
The graph-based Pixie method, which uses the graph of pin-to-board relations, correctly understands that the category of query is ``plants'' and it recommends items in that general category. However, it does not find the most relevant items.
Combining both visual/textual and graph information, \name~ is able to find relevant items that are both visually and topically similar to the query item.

\begin{figure}[t]
\centering
\includegraphics[width=0.45\textwidth]{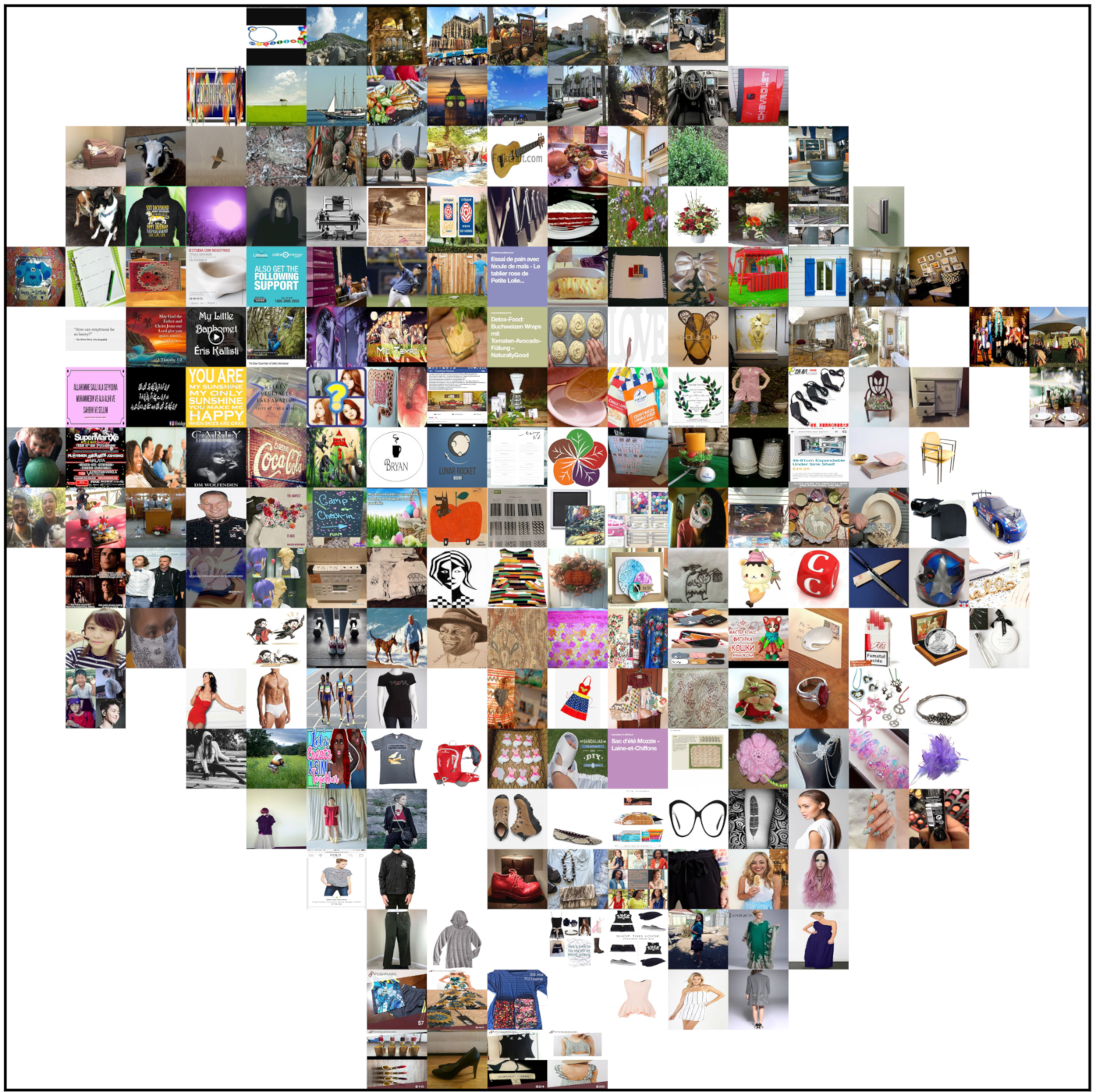}
\caption{t-SNE plot of item embeddings in 2 dimensions.}
\label{fig:tsne}
\vspace{-2mm}
\end{figure}

In addition, we visualize the embedding space by randomly choosing $1000$ items and compute the 2D t-SNE coordinates from the \name embedding, as shown in Figure \ref{fig:tsne}.\footnote{Some items are overlapped and are not visible.} We observe that the proximity of the item embeddings corresponds well with the similarity of content, and that items of the same category are embedded into the same part of the space. Note that items that are visually different but have the same theme are also close to each other in the embedding space, as seen by the items depicting different fashion-related items on the bottom side of the plot.

\begin{table}[t]
	\begin{tabular}{c | c | c | c | c}
	\hline
	Methods & Win & Lose & Draw & Fraction of wins  \\
	\hline
	\name~ vs. Visual  & 28.4\% & 21.9\% & 49.7\% & 56.5\%  \\
	\name~ vs. Annot.  & 36.9\% & 14.0\%  & 49.1\% & 72.5\%\\
	\name~ vs. Combined & 22.6\% & 15.1\% & 57.5\% & 60.0\% \\
	\name~ vs. Pixie & 32.5\% & 19.6\% & 46.4\% & 62.4\% \\
	\hline
	\end{tabular}
	\caption{Head-to-head comparison of which image is more relevant to the recommended query image. 
	}
	\label{fig:user_study_head2head}
	\vspace{-5mm}
\end{table}

\subsection{Production A/B Test}
\label{sec:ab_test}
Lastly, we also report on the production A/B test experiments, which compared the performance of \name to other deep learning content-based recommender systems at Pinterest on the task of homefeed recommendations.
We evaluate the performance by observing the lift in user engagement. 
The metric of interest is \emph{repin rate}, which measures the percentage of homefeed recommendations that have been saved by the users. A user saving a pin to a board is a high-value action that signifies deep engagement of the user. It means that a given pin presented to a user at a given time was relevant enough for the user to save that pin to one of their boards so that they can retrieve it later.

\cut{
Table \ref{tab:ab_test} presents the results for the homefeed A/B test which compares the homefeed recommendation made by
\name \ embeddings with that of the two currently deployed recommendation engines based on the Annotation
and Pixie baselines. 
The A/B test was run over $7$ days and involved over 300,000 users.
\begin{table}[h]
	\begin{tabular}{c | c | c | c} 
	\hline
	Baselines & \name & Annotation & Pixie \\
	\hline
	Repin rate & ??? & ?? & ??? \\
	\hline
	\end{tabular}
	\caption{Homefeed A/B test results.  }
	\label{tab:ab_test}
	\vspace{-8mm}
\end{table}
} 

We find that \name consistently recommends pins that are
more likely to be re-pinned by the user than the alternative methods. Depending on the particular setting, we observe 10-30\% improvements in repin rate over the Annotation and Visual embedding based recommendations.

\subsection{Training and Inference Runtime Analysis}
\label{sec:runtime}

One advantage of \framework{s} is that they can be made inductive \cite{hamilton2017representation}:
at the inference (\ie, embedding generation) step, we are able to compute embeddings for items that were not in the training set. 
This allows us to train on a subgraph to obtain model parameters, and then make embed nodes that have not been observed during training. Also note that it is easy to compute embeddings of new nodes that get added into the graph over time. This means that  recommendations can be made on the full (and constantly growing) graph.
Experiments on development data demonstrated that training on a subgraph containing $300$ million
items could achieve the best performance in terms of hit-rate (\ie, further increases in the training set size did not seem to help), reducing the runtime by a factor of $6$ compared to training on the full graph. 

Table \ref{fig:runtime} shows the the effect of batch size of the minibatch SGD on the runtime of \name training procedure, using the mean-pooling-hard variant. For varying batch sizes, the table shows: (1) the computation time, in milliseconds, for each minibatch, when varying batch size; (2) the number of iterations needed for the model to converge; and (3) the total estimated time for the training procedure.
Experiments show that a batch size of $2048$ makes training most efficient. 

When training the \name variant with importance pooling, another trade-off comes from choosing the size of neighborhood $T$.
Table \ref{fig:runtime} shows the runtime and performance of \name when $T=10$, $20$ and $50$.
We observe a diminishing return as $T$ increases, and find that a two-layer GCN with neighborhood size $50$ can best capture the neighborhood information of nodes, while still being computationally efficient.

\begin{table}[t]
	\begin{tabular}{c | c | c | c} 
	\hline
	Batch size & Per iteration (ms) & \# iterations & Total time (h) \\
	\hline
	512 & 590 & 390k & 63.9 \\
	1024 &  870& 220k & 53.2 \\
	2048 & 1350  & 130k & 48.8 \\
	4096 & 2240 & 100k & 68.4 \\
	\hline
	\end{tabular}
	\caption{Runtime comparisons for different batch sizes.}
	\label{fig:runtime}
	\vspace{-5mm}
\end{table}

\begin{table}[t]
  \begin{tabular}{c | c | c | c}
    \hline
    \# neighbors & Hit-rate & MRR & Training time (h) \\
    \hline
    10 & 60\% & 0.51 & 20 \\
    20 & 63\% & 0.54 & 33 \\
    50 & 67\% & 0.59 & 78 \\
    \hline
  \end{tabular}
  \caption{Performance tradeoffs for importance pooling.}
  \label{fig:runtime_importance_pooling}
  \vspace{-5mm}
\end{table}

After training completes, due to the highly efficient MapReduce inference pipeline, the whole inference procedure to generate embeddings for $3$ billion items can finish in less than 24 hours.

\section{Conclusion}
\label{sec:conclusion}
We proposed \name, a random-walk graph convolutional network (GCN). \name is a highly-scalable GCN algorithm capable of learning embeddings for nodes in web-scale graphs containing billions of objects. 
In addition to new techniques that ensure scalability, we introduced the use of importance pooling and curriculum training that drastically improved embedding performance.
We deployed \name at Pinterest and comprehensively evaluated the quality of the learned embeddings on a number of recommendation tasks, with offline metrics, user studies and A/B tests all demonstrating a substantial improvement in recommendation performance. 
Our work demonstrates the impact that graph convolutional methods can have in a production recommender system, and we believe that \name can be further extended in the future to tackle other graph representation learning problems at large scale, including knowledge graph reasoning and graph clustering.

\subsection*{Acknowledgments}
The authors acknowledge Raymond Hsu, Andrei Curelea and Ali Altaf for performing various A/B tests in production system,
Jerry Zitao Liu for providing data used by Pixie\cite{pixie2018}, and Vitaliy Kulikov for help in nearest neighbor query of the item embeddings.

\bibliographystyle{ACM-Reference-Format}
\bibliography{refs} 

\end{document}